\documentclass{PoS}

\usepackage{mathtools}
\usepackage{amsfonts}
\usepackage{amssymb}
\usepackage{fdsymbol}
\usepackage{float}
\newcommand{\sgn}{\textrm{sgn}}
\newcommand{\tr}{\textrm{tr}}

\usepackage{caption}
\usepackage{subcaption}

\title{Two-colour QCD at finite density with two flavours of staggered quarks}

\ShortTitle{LQC$_2$D at $\mu\neq 0$ with $N_f=2$ staggered quarks}

\author{\speaker{Lukas Holicki}\thanks{This work was supported by the Helmholtz International Center for FAIR within the LOEWE initiative of the State of Hesse.}\\
	Institut f\"ur Theoretische Physik, Justus-Liebig-Universit\"at Gie\ss{}en, 35392 Gie\ss{}en, Germany\\
	E-mail: \email{Lukas.Holicki@physik.uni-giessen.de}
}

\author{Jonas Wilhelm\\
	Institut f\"ur Theoretische Physik, Justus-Liebig-Universit\"at Gie\ss{}en, 35392 Gie\ss{}en, Germany\\
	E-mail: \email{Jonas.Wilhelm@physik.uni-giessen.de}
}

\author{Dominik Smith\\
	Institut f\"ur Theoretische Physik, Justus-Liebig-Universit\"at Gie\ss{}en, 35392 Gie\ss{}en, Germany\\
	E-mail: \email{Dominik.D.Smith@theo.physik.uni-giessen.de}
}

\author{Bj\"orn Wellegehausen\\
	Institut f\"ur Theoretische Physik, Justus-Liebig-Universit\"at Gie\ss{}en, 35392 Gie\ss{}en, Germany\\
	E-mail: \email{Bjoern.Wellegehausen@theo.physik.uni-giessen.de}
}

\author{Lorenz von Smekal\\
	Institut f\"ur Theoretische Physik, Justus-Liebig-Universit\"at Gie\ss{}en, 35392 Gie\ss{}en, Germany\\
	E-mail: \email{Lorenz.Smekal@theo.physik.uni-giessen.de}
}

\abstract{In this contribution we revisit simulations of two-color QCD
  with rooted staggered quarks at finite density, where
  baryon-number spontaneously breaks and a diquark
  condensate forms. We thereby pay special attention to simulating
  outside the lattice-artifact bulk phase, in which $Z_2$ monopoles
  condense, and investigate some of the consequences of this, e.g. on
  the chiral and the diquark condensate which were known to be well described by
  chiral effective field theory. Not surprisingly, on finer lattices
  outside the bulk phase the quark condensate now requires additive
  renormalization before it can be compared with effective field
  theory predictions. The subtraction must necessarily depend on the chemical
  potential, however. The diquark condensate is not affected by this
  problem and remains in good agreement with these 
  predictions.
  We also compare staggered with Wilson quarks to demonstrate that the
  two fermion discretizations yield qualitatively different results
  well below half-filling already. We close with prelimiary results
  for the Goldstone spectrum to demonstrate that the continuum pattern
  is recovered also with staggered quarks outside the bulk phase.}

\FullConference{34th annual International Symposium on Lattice Field Theory\\
		24-30 July 2016\\
		University of Southampton, UK}

\begin{document}

\section{Introduction}
\label{sec:intro}

Lattice simulations of two-colour QCD (QC$_2$D) can be performed at
finite density without a sign problem and by now have a long history
already \cite{Hands:1999md,Kogut:2001na}.
The physics of the bosonic diquark baryons is believed to be fairly
well understood and qualitatively resembles QCD at finite isospin
density with pion condensation \cite{Son:2000xc}. There is good
guidance from effective field theory and random matrix theory predictions
\cite{Kogut:2000ek,Kogut:2003ju,Kanazawa:2011tt} and model studies
of the BEC-BCS crossover inside the condensed phase
\cite{Strodthoff:2011tz,Kamikado:2012bt}.
In QC$_2$D diquarks play a dual role as two-color baryons and
pseudo-Goldstone bosons of the dynamical breaking of an extended
chiral symmetry. When they condense they are expected to form a
superfluid which changes in nature from a Bose-Einstein condensate of
tightly bound diquarks to a BCS-like pairing of quarks as chiral symmetry gets
gradually restored with increasing density. Recent results with both,
staggered \cite{Braguta:2016cpw} and Wilson \cite{Boz:2015ppa}  quarks
have confirmed previous evidence from coarser lattices for this
superfluid phase. In line with the predictions from  chiral effective
Lagrangians, at zero temperature it first occurs when the quark
chemical potential $\mu$ reaches half the pion mass, and this is also
correctly reproduced in effective lattice theories for heavy quarks
\cite{Scior:2015vra}, from the same combined strong-coupling and hopping
expansion techniques that are actively being developed for cold and
dense QCD \cite{Philipsen:2016wjt}.

Here, we revisit the low-temperature part of the QC$_2$D phase diagram
with rooted staggered quarks. After reproducing previous results for
the chiral- and diquark-condensates, the quark-number density and the Goldstone
spectrum along the $\mu$ direction at $\beta=1.5$, we assess the
influence of the lattice-artifact bulk phase dominated by
$Z_2$ monopoles. With an improved gauge action and a somewhat larger
lattice coupling of $\beta=1.7$ on the other hand, these monopoles are
sufficiently suppressed to connect with continuum results. While this
then requires a $\mu$-dependent additive renormalization of the
chiral condensate, the extrapolated diquark condensate and
quark-number density near the onset transition at half the pion mass
remain in very good agreement with the effective field theory
predictions. Beyond that one might speculate to see evidence of the
BEC-BCS crossover at larger $\mu$, but this gets hard to disentangle
from discretization artifacts at larger $\mu$ which we assess by
comparing staggered with Wilson fermions revealing significant
qualitative differences between the two fermion discretizations
already well below half-filling.

Finally but maybe most importantly our prelimiary Goldstone spectrum
near the onset transition provides good evidence that the symmetry
breaking pattern of QC$_2$D in the continuum is recovered also with
staggered quarks outside the bulk phase and hence in the continuum
limit, despite the wrong antiunitary symmetries of staggered
fermions at finite lattice spacing \cite{Kogut:2000ek,Kogut:2003ju}.

\section{General Setup}\label{sec:setup}

To study the spontaneous breaking of baryon number on a finite
lattice, one adds the diquark source term $\sim \lambda$ to the
Lagrangian as an  explicitly symmetry-breaking external field. The
staggered fermion action of QC$_2$D is then expressed in a
Nambu-Gorkov basis as \cite{Kogut:2001na}
	\begin{equation}
		S_f = \overline \chi D(\mu) \chi + \frac \lambda 2 \left( \chi^T\tau_2\chi+\overline\chi\tau_2\overline\chi^T \right) = \frac 1 2 \left( \overline\chi, \chi^T \tau_2 \right) \underbrace{ \begin{pmatrix} \lambda & D(\mu) \\ -D^\dagger(\mu) & \lambda \end{pmatrix} }_{=: A} \begin{pmatrix} \tau_2 \overline \chi^T \\ \chi \end{pmatrix},
		\label{eq:diquark_action}
	\end{equation}
where $D$ denotes the staggered Dirac operator, such that $\det\, A =
\det \left( D^\dag(\mu)D(\mu) + \lambda^2 \right)$.
Results are extrapolated to $\lambda\rightarrow 0$, and the diquark
condensate is obtained from
\begin{equation}
	\langle qq \rangle = \langle \chi^T \tau_2 \chi \rangle \propto \frac{\partial \ln Z}{\partial \lambda} \bigg|_{\lambda\rightarrow0} .
\end{equation}
We use the RHMC algorithm \cite{Clark2005} to simulate a rooted
fermion determinant $\det A^{N_f/8}$.

The prediction from the leading order chiral Lagrangian with diquark
source $\lambda $  for fundamental staggered
quarks in QC$_2$D is that the vaccum rotates at fixed
\text{$\langle\overline q  q\rangle^2+\langle q q \rangle^2 $} from a
chiral to a diquark condensate with $\tan\phi=\lambda/ m$ and rotation angle
$\alpha(\mu)$ as $\mu$ increases, from $\alpha = \phi $ for $\mu=0$
to $\alpha = \pi/2$ for $\mu \gg \mu_c = m_\pi/2$, which is obtained
from \cite{Kogut:2000ek}
	\begin{equation}
		\begin{array}{ccc}
			4\mu^2 \cos\alpha \sin\alpha = m_\pi^2
                        \sin(\alpha-\phi)\, , \hskip .1cm &  \hskip .2cm
                        \mbox{such that}  \hskip .2cm
                        &  n_B = 8 N_f F^2 \mu \sin^2 \alpha  \, ,\\
			\langle \overline q q \rangle = 2 N_f G \cos
                        \alpha\, ,  &
                        \mbox{and} & \langle qq \rangle = 2 N_f G \sin
                        \alpha \, .\\
		\end{array}
		\label{eq:xpt_condensates}
	\end{equation}
We fit these equations to the results from our simulations. For
$\lambda\to 0$ they describe the zero-temperature onset of diquark
condensation at $\mu_c = m_\pi/2$, which is determined as a fit
parameter here and compared with an independent
determination of $m_\pi$ from spectroscopy as a check.

%

\section{Results in the bulk phase}\label{subsec:bulkphase}

After reproducing the results from \cite{Kogut:2003ju} for $\langle qq \rangle$, $\langle\overline q q \rangle$ and the quark number-density $\langle n \rangle$
with $N_f=4$ at $\beta=1.5$, $N_s=12$, $N_t=24$, $am=0.025$, $a\lambda=0.0025$ as a check, we have performed the analogous simulations with the fourth root for $N_f=2$, see Fig.~\ref{fig:xpt_fit}. The corresponding Goldstone spectrum is shown in Fig.~\ref{fig:goldstone_spectrum_beta1.5}. As for the $N_f = 4$ results from \cite{Kogut:2003ju} it reflects the symmetry breaking pattern of fundamental staggered quarks (shown here as $\chi$PT fit) which resembles that of adjoint
QCD or $G_2$-QCD in the continuum as most notably seen in the behavior
of the pion branch above the onset that here results at $a\mu_c =
0.18889(45)$ from the fit. Within the errors this well agrees with our
spectroscopic result from the pion correlator at $\mu = 0$ which yields
$am_\pi/2 = 0.1887(06)$.

\hspace{-5mm}
\begin{minipage}[c]{0.47\linewidth}
\begin{figure}[H]
		\centering
		\includegraphics[width=\textwidth]{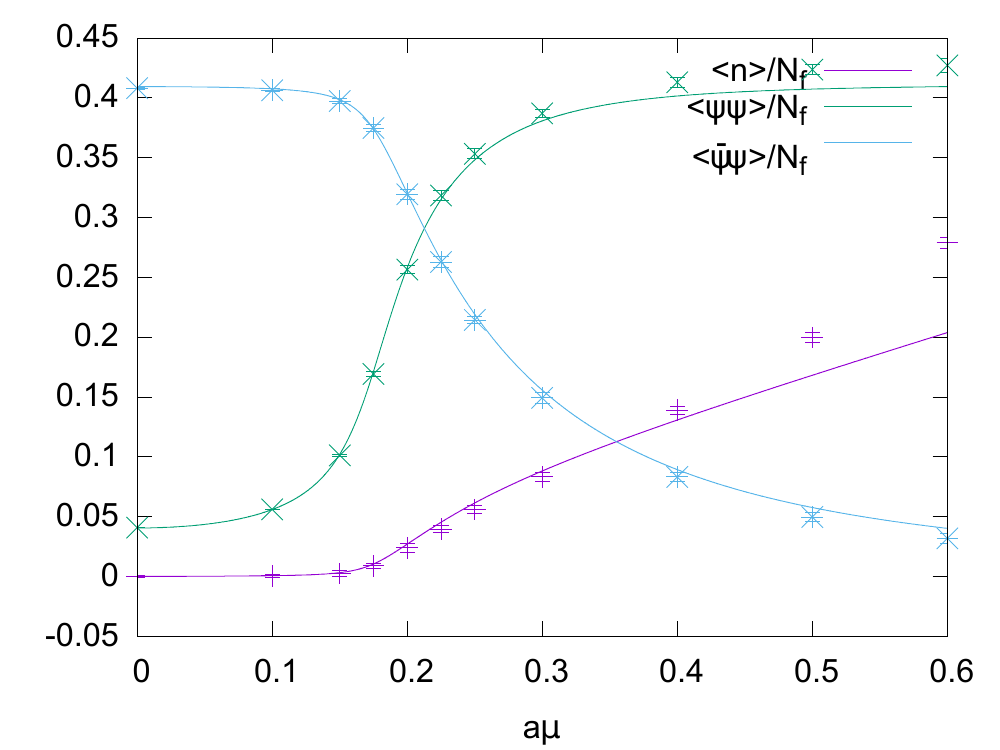}
		\caption{Fit of lattice data to $\chi$PT predictions
                  of condensates and density at $\beta = 1.5$.}
		\label{fig:xpt_fit}
	\end{figure}
\end{minipage}\hspace{5mm}
\begin{minipage}[c]{0.47\linewidth}
	\begin{figure}[H]
		\centering
		\includegraphics[width=\textwidth]{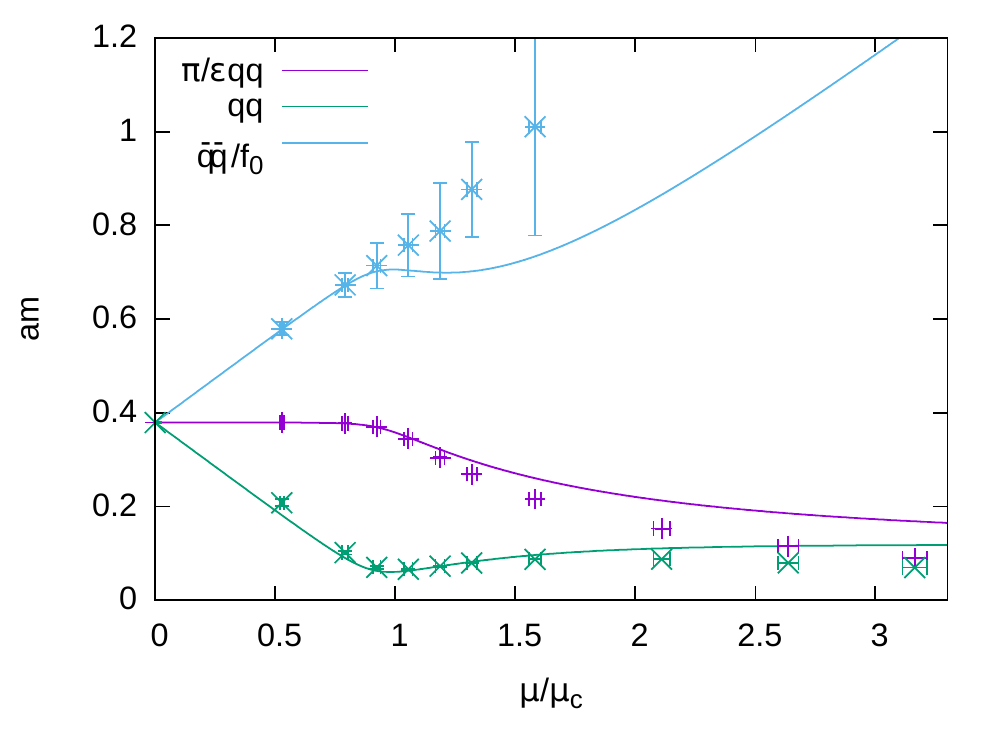}
		\caption{Goldstone spectrum for 
                  Fig.~\protect\ref{fig:xpt_fit}
                  without disconnected contributions in $\bar q \bar q/f_0$.}
		\label{fig:goldstone_spectrum_beta1.5}
	\end{figure}
\end{minipage}\vspace{2mm}

The scalar $f_0$ mixes with the heavy (anti)diquark mode when baryon number is
broken in the diquark-condensation phase. The discrepancy of the
corresponding $\bar q \bar q /f_0$ branch with $\chi$PT there is to
a large extend due to disconnected contributions to $f_0$ which we did
not calculate here.

A potential problem in these simulations is the lattice-artifact
bulk phase of SU(2) with $Z_2$-monopole condensation at strong coupling
\cite{Halliday:1981tm}. We have therefore
measured the $Z_2$-monopole density $z = 1 - \frac{1}{N_c}  \sum_{c}
\prod_{P\in\partial c} \sgn(\tr \, P)$ sensitive to preferred signs of
plaquettes on the faces of elementary cubes $c$. Its dependence on the
lattice coupling with and without quarks is shown in
Fig.~\ref{fig:bulk_phase_dependencies}. It is rather insensitive to
unquenching, at $\beta=1.5$ even with $am = 0.01$ (and $\mu=0$) we obtain
$\langle z \rangle = 0.8840(1)$ which shows that this is well below
the crossover and deep inside the bulk phase.

To suppress $Z_2$ monopoles we therefore use a tree-level
Symanzik-improved gauge action \cite{Weisz:1983bn},
cf.~Fig.~\ref{fig:bulk_phase_dependencies}. From extended meson spectroscopy
 \cite{SchefflerDiss2015} we conclude that the best compromise between
 discretization and finite-volume effects with the improved action
 on our present $16^3\times 32$ lattice is obtained for $\beta = 1.7$,
 see Fig.~\ref{fig:meson_masses_over_beta}.
 The $Z_2$-monopole density 
 is then down to $\langle z \rangle = 0.2734(7)$.

\hspace{-5mm}
\begin{minipage}[c]{0.48\linewidth}
	\begin{figure}[H]
		\centering
		\includegraphics[width=\textwidth]{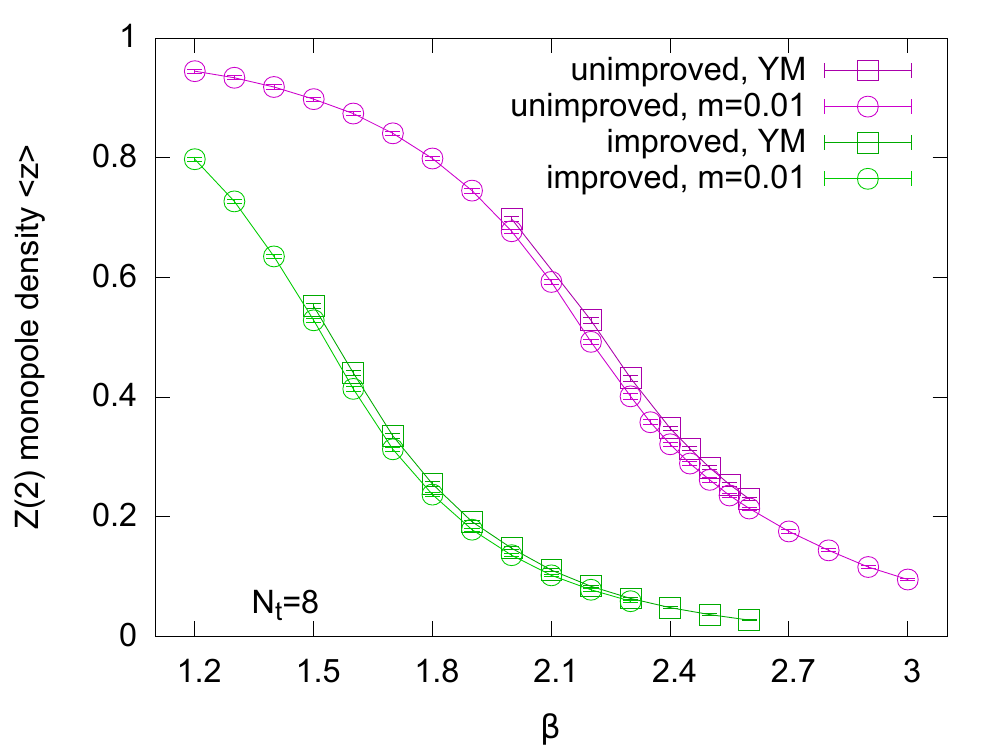}
		\caption{$Z_2$-monopole density, $N_s=12$, $N_t=8$.}
		\label{fig:bulk_phase_dependencies}
	\end{figure}
\end{minipage}\hspace{5mm}
\begin{minipage}[c]{0.48\linewidth}
	\begin{figure}[H]
		\centering
		\includegraphics[width=\textwidth]{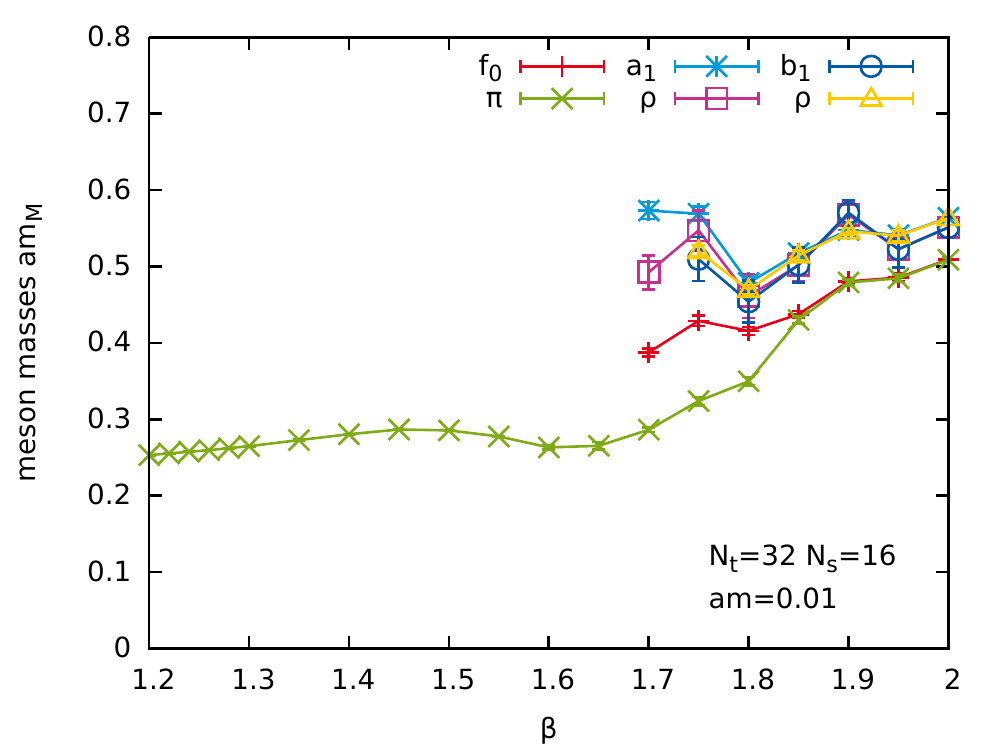}
		\caption{Meson masses over $\beta$ from \protect\cite{SchefflerDiss2015}.}
		\label{fig:meson_masses_over_beta}
	\end{figure}
\end{minipage}

\section{Chiral and diquark condensates outside the bulk phase}

\begin{figure}[!ht]
	\begin{subfigure}[c]{0.47\textwidth}
		\centering
		\includegraphics[width=\textwidth]{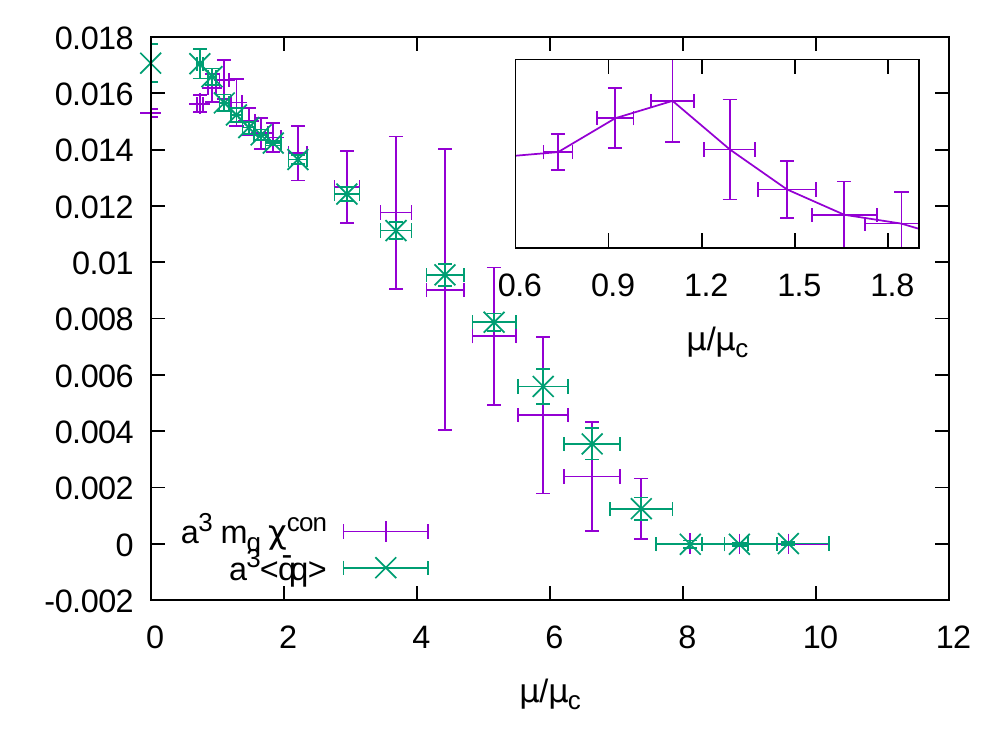}
		\subcaption{$\langle \overline q q \rangle$ and
                  $\chi^\textmd{con}$ with a zoom to its singular part.}
		\label{fig:chiral_condensate_renorm}
	\end{subfigure}
	\begin{subfigure}[c]{0.47\textwidth}
		\centering
		\includegraphics[width=\textwidth]{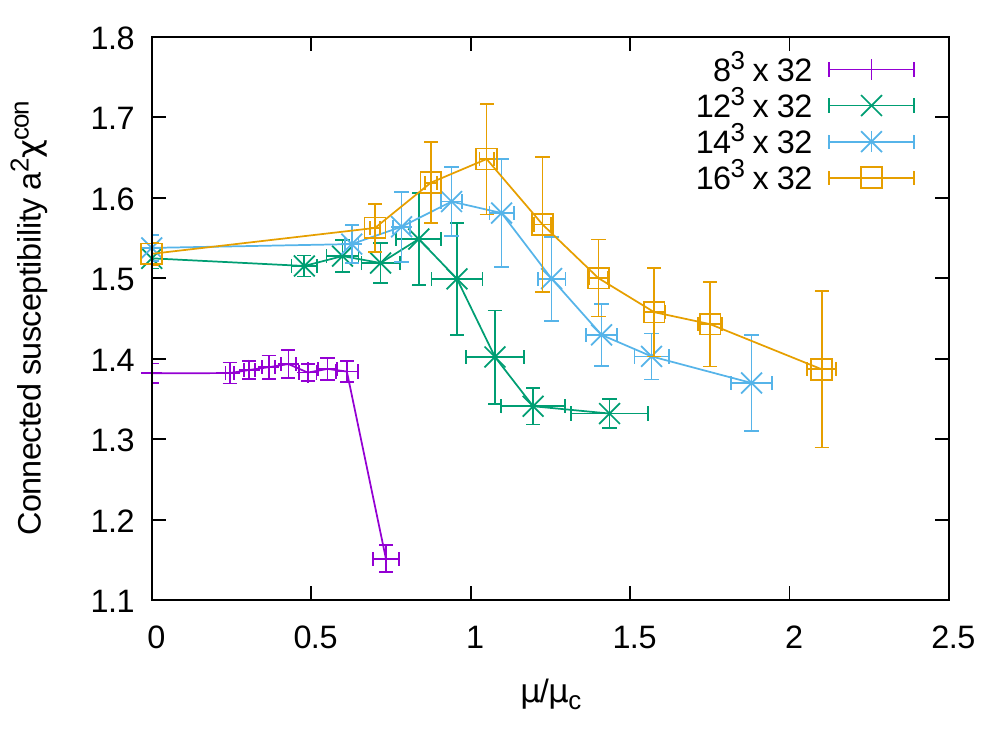}
		\subcaption{volume dependence of $\chi^\textmd{con}\!$.}
		\label{fig:chiral_con_susc_vol_dep}
	\end{subfigure}
	\caption{$\langle\overline qq\rangle$ and its connected
          susceptibility ($16^3\times 32$ lattice,
          \text{$am=0.01$}, $\beta=1.7$, $a\lambda=0.005$).}
\end{figure}

With the improved gauge action, $\beta=1.7$ and $am=0.01$ on the
$16^3\times32$ lattice, where ${m_\pi}/{m_\rho}=0.5816(27)$, we have
then performed a low temperature scan of the phase diagram. The
chiral condensate as a function of the quark chemical potential in
units of $\mu_c$ is shown in
Fig.~\ref{fig:chiral_condensate_renorm}. Not surprisingly, on the
finer lattice it now requires additive renormalization. In finite
temperature simulations at $\mu =0$ this can be achieved with a
\textit{connected susceptibility subtraction}, \text{$\Sigma
  = \langle\overline qq\rangle_{m_q} - m_q \chi^\textmd{con}$}, by
observing that (up to a factor $m_q$) chiral condensate and susceptibility
share the same quadratic divergence $c_\mathrm{UV}/a^2$ and that this
mainly comes from the connected part  $\chi^\textmd{con}$ of the
latter \cite{Unger2010}.
To see whether this subtraction can be used at finite $\mu $ as well,
we also included our results for  $\chi^\textmd{con}$ in
Fig.~\ref{fig:chiral_condensate_renorm}.  We observe that they both
drop to zero above $\mu_c$, so the quadratic divergence must depend on
$\mu $ in the diquark condensation phase,\footnote{As it
 also does for free lattice fermions of mass $m$ at $\mu > m$.}
i.e.~$c_\mathrm{UV} = c_\mathrm{UV}(\mu)$. Secondly, and more
importantly, however, we also observe evidence of a singular
contribution in the connected susceptibility $\chi^\textmd{con}$
near the diquark-condensation transition at $\mu=\mu_c$, as seen from
its volume dependence in Fig.~\ref{fig:chiral_con_susc_vol_dep}. This
is different from the chiral transition where the thermodynamic
singularity resides in the disconnected part of the
susceptibility. It means that this singularity in $\chi^\textmd{con}$
will dominate over $c_\mathrm{UV}/a^2$ at finite $a$ in the infinite
volume limit. The chiral condensate on the other hand, at zero
temperature, must remain independent of $\mu$ for
$\mu<\mu_c$. It does not have such a singular contribution
and it would be unphysical to introduce one with the connected
susceptibility subtraction. At any rate, this would introduce a
$\mu$-dependence below $\mu_c$ and hence a Silver-Blaze problem.
Therefore, a different ($\mu$-dependent) subtraction of
the chiral condensate is required at finite density.

\begin{figure}[!ht]
			\begin{subfigure}[c]{0.47\textwidth}
				\centering
				\includegraphics[width=\textwidth]{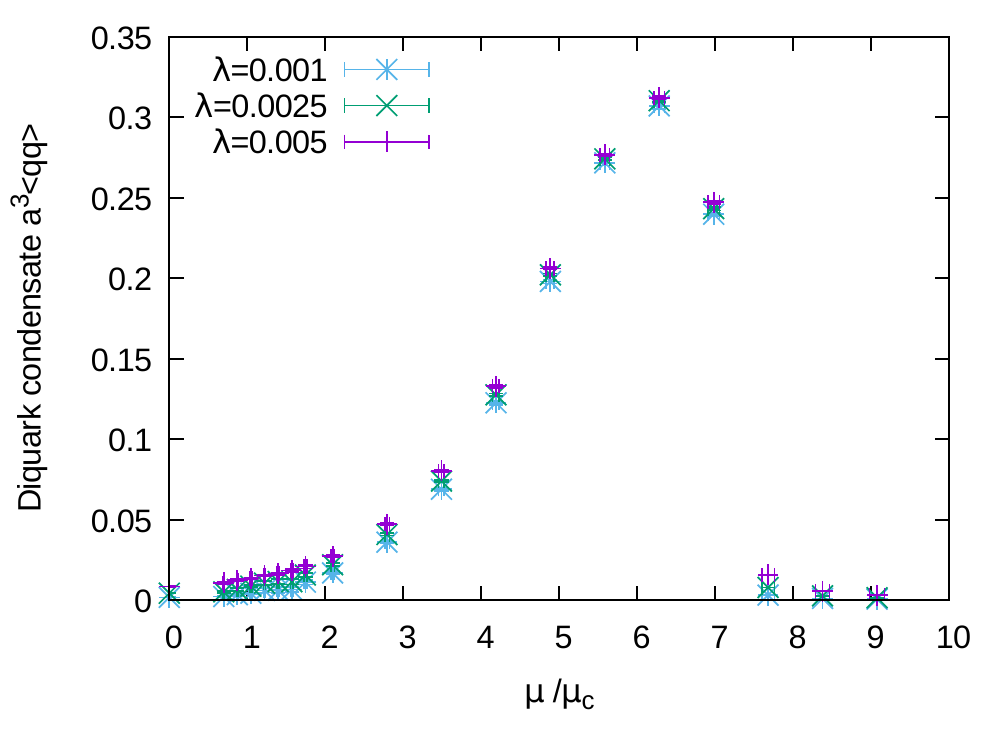}
				\subcaption{$\mu$-dependence of
                                  $\langle qq\rangle$
                                   for various $\lambda$.}
				\label{fig:diquark}
			\end{subfigure}
			\begin{subfigure}[c]{0.47\textwidth}
				\centering
				\includegraphics[width=\textwidth]{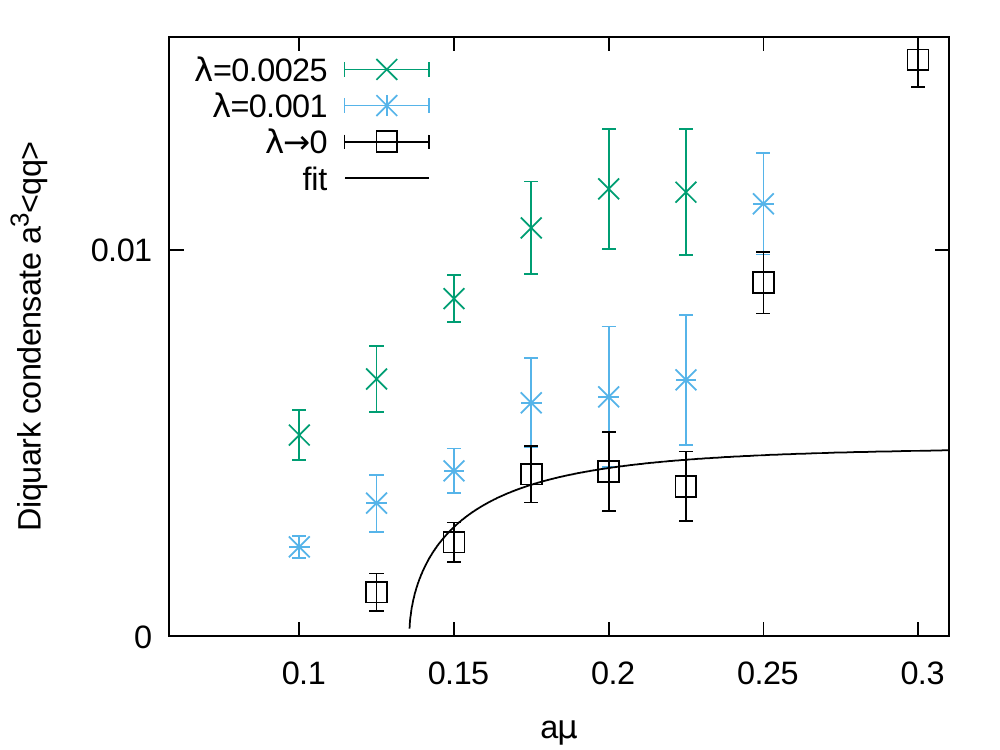}
				\subcaption{$\lambda\rightarrow0$
                                  extrapolation of $\langle qq\rangle$
                                  with  $\chi$PT fit.}
				\label{fig:diquark_zoom}
			\end{subfigure}
			\caption{$\langle qq\rangle$ and leading-order
                          $\chi$PT fit
                          $\propto \sqrt{ 1 - (\mu_c/\mu)^4} $  ($
                          16^3\times32$ lattice, $am=0.01$, $\beta=1.7$).}
		\end{figure}

The diquark condensate in Fig.~\ref{fig:diquark} is unaffected by this
problem.  Its $\lambda\rightarrow 0$ extrapolation now yields the
diquark-condensation transition at $a\mu_c = 0.1356(86)$ from the fit
to the $\chi$PT prediction shown in Fig.~\ref{fig:diquark_zoom}. This
is again consistent with the mass extracted from the pion correlator at
$\mu =0$ giving $am_\pi/2=0.1428(26)$ here for comparison. The sudden
increase of the diquark condensate as compared to the $\chi$PT
fit at around $1.6 \mu_c $ was also observed in \cite{Braguta:2016cpw}
and interpreted as evidence of the BEC-BCS crossover. While this also
agrees with the estimates from model predictions for
the crossover region around $\mu \sim 0.8 \, m_\pi$
\cite{Strodthoff:2011tz,Kamikado:2012bt}, we note however,
that it might already be difficult to disentangle from the discretization
artifacts setting in at larger $\mu$ as we will discuss below.
The sharp drop of $\langle qq \rangle$ at $\mu$ around $ 7 \mu_c $ is close
to half filling, where the quark-number density in lattice units
reaches half its saturation value of maximum fermion occupancy on the
finite lattice. Both condensates vanish at
lattice saturation and other observables approach their quenched values.

\vspace{-7mm}

\section{Staggered versus Wilson fermions at finite density}

\begin{figure}[t]
\vspace*{-.4cm}

		\begin{subfigure}[c]{0.47\textwidth}
		  \includegraphics[width=\textwidth]{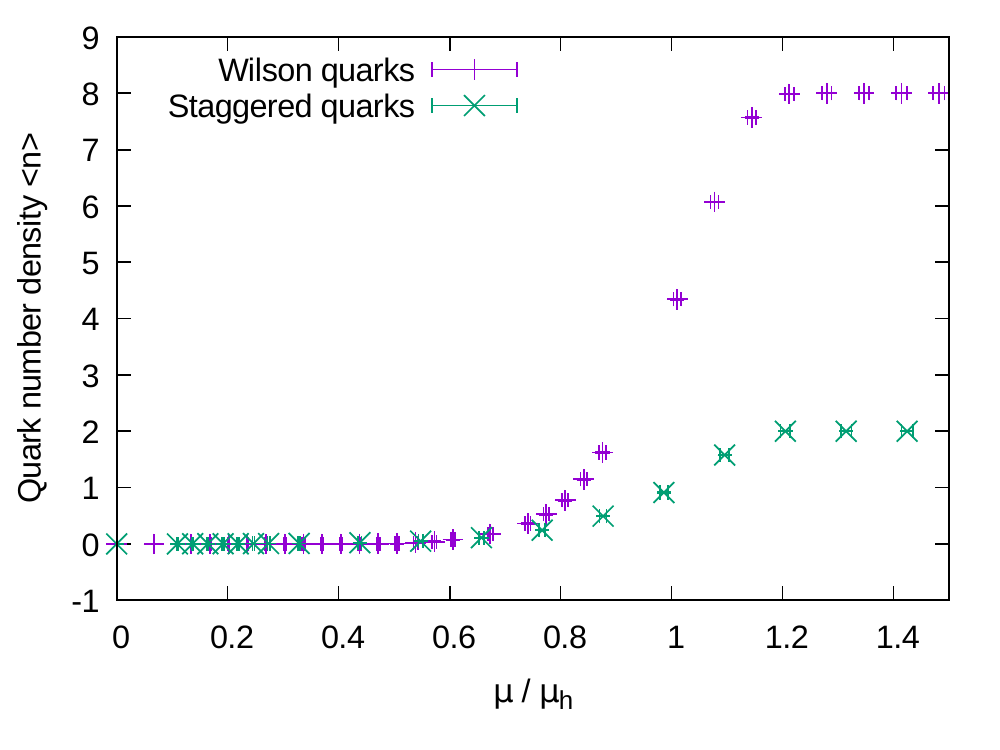}

                                  \vspace*{-.2cm}
			\subcaption{quark-number density in lattice units.}
			\label{fig:wilson_stag_comp_qdens}
		\end{subfigure}
		\begin{subfigure}[c]{0.47\textwidth}
		  \includegraphics[width=\textwidth]{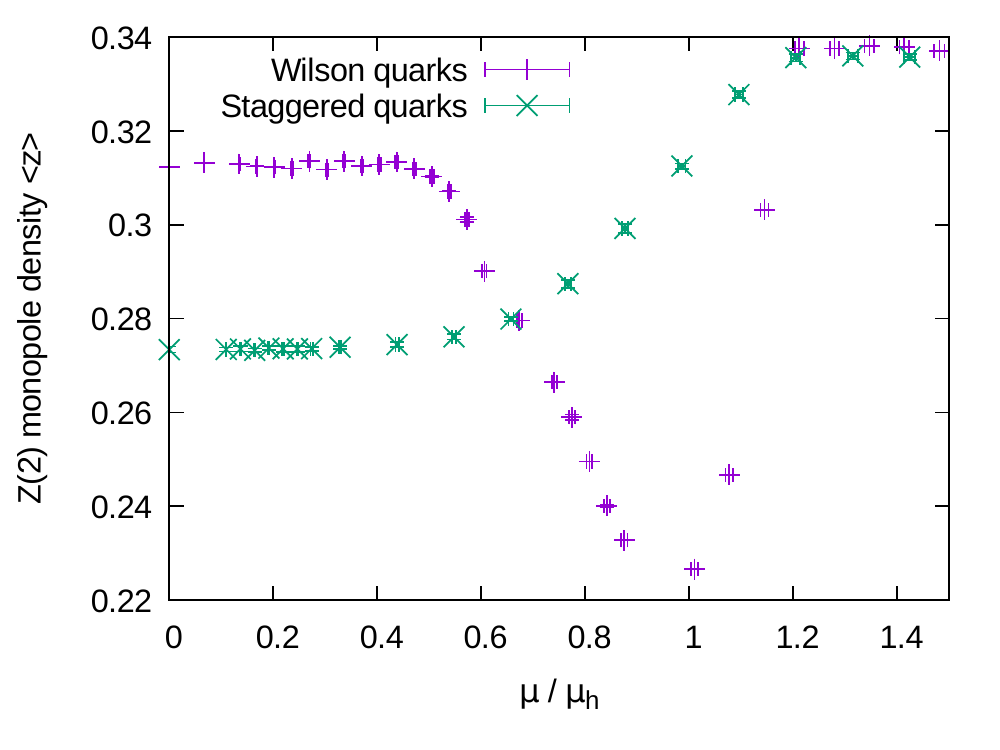}

                                                          \vspace*{-.2cm}
			\subcaption{$\mu$-dependence of $Z_2$-monopole
                          density.}
			\label{fig:wilson_stag_comp_z2mono}
		\end{subfigure}

                \vspace*{-.2cm}
	\caption{Comparison of two-flavor simulations from fourth-root
          staggered ($16^3\times32$
          lattice, $am=0.01$) and Wilson quarks ($10^3\times20$
          lattice, $\kappa=0.124689$) at finite density with
          $\beta=1.7$, $a\lambda=0.001$.}

        \vspace*{-.4cm}
\end{figure}

The maximum number of fermions on a finite lattice depends on the
fermion discretization, of course. In lattice units, the saturation
density for Wilson fermions is $a^3 n_\mathrm{sat}^\mathrm{W} = 2 N_f N_c =
8$ in our two-flavor QC$_2$D simulations, while for staggered fermions
$a^3 n_\mathrm{sat}^\mathrm{KS} = N_c = 2$ here, independent of
rooting. In order to compare the two discretizations without detailed
mapping of the different scales we plot the respective quark-number
densities over the chemical potential in units of its value
$\mu_\mathrm{h}$ at half filling, where $n = n_\mathrm{sat}/2$ in
Fig.~\ref{fig:wilson_stag_comp_qdens}. When they
start to deviate, this can be taken as a signal of strong
discretization effects at least in one of the two.

As an example of observables which approache their quenched values
with lattice saturation we plot the $Z_2$-monopole densities of
both fermion discretizations in
Fig.~\ref{fig:wilson_stag_comp_z2mono}.
They start to show a notably different qualitative behavior at $\mu $
well below $\mu_\mathrm{h}$ already,  before they approach saturation
where their values are both consistent with the quenched result.
The same is true for the Polyakov loop. It remains essentially
independent of the chemical potential for staggered fermions as
already observed in \cite{Braguta:2016cpw}. In contrast, it is well
known from many previous studies of QCD-like theories with Wilson
quarks \cite{Boz:2015ppa,Maas:2012wr,Wellegehausen:2013cya}, and the
effective lattice theories for heavy quarks
\cite{Scior:2015vra,Philipsen:2016wjt}, that the Polyakov loop starts
to rise in the dense phase with a peak around half filling as also observed in
Fig.~\ref{fig:wilson_stag_comp_ploop} for our simulations with Wilson
quarks on a $10^3\times20$ lattice at $\beta=1.7$, $\kappa=0.124689$.

\vspace*{-.2cm}

\hspace{-5mm}
	\begin{minipage}[c]{0.46\textwidth}
		\begin{figure}[H]
			\centering
			\includegraphics[width=\textwidth]{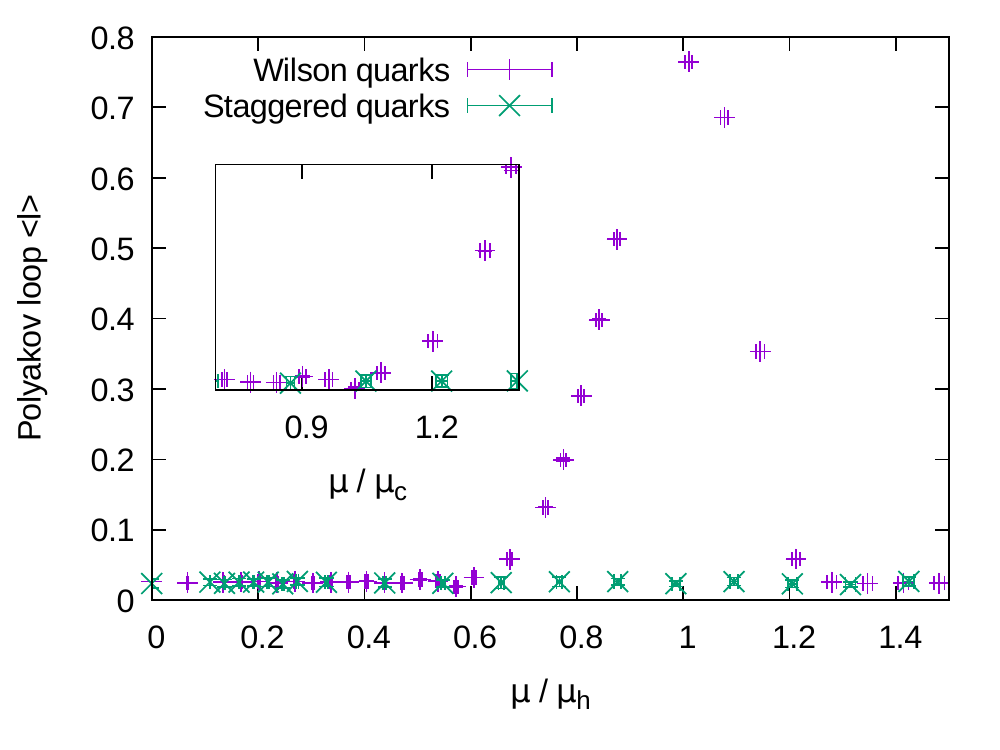}

                        \vspace*{-.5cm}
			\caption{Polyakov loop with Wilson and staggered
                          quarks, overview and inlay near
                          $\mu_c$.}
			\label{fig:wilson_stag_comp_ploop}
		\end{figure}
	\end{minipage}\hspace{5mm}
	\begin{minipage}[c]{0.46\textwidth}
		\begin{figure}[H]
			\centering
			\includegraphics[width=\textwidth]{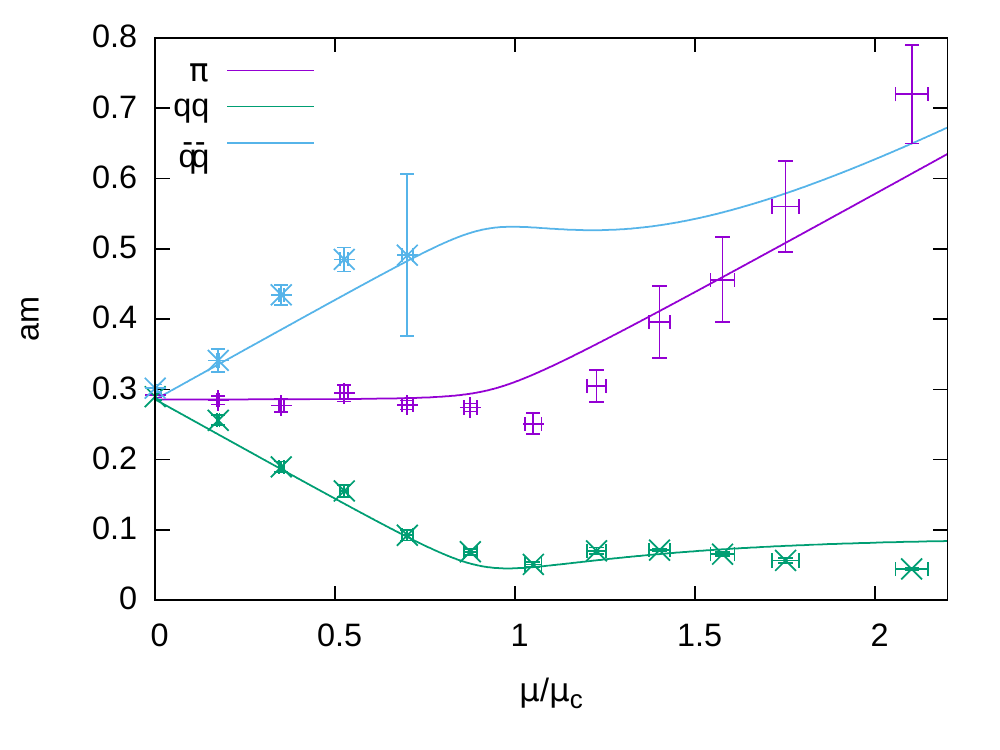}

                        \vspace*{-.5cm}
			\caption{Goldstone spectrum outside bulk phase
                          with continuum prediction from $\chi$PT.}
			\label{fig:gsmb1.7}
		\end{figure}
	\end{minipage}


\section{Summary and outlook}

We have simulated two-color QCD with fourth-root staggered quarks
for two flavors at finite density. We have thereby compared results
from lattice parameters employed previously at $\beta = 1.5$, deep
inside the bulk phase with high $Z_2$-monopole density, with
results from an improved gauge action together with a somewhat larger lattice
coupling of $\beta=1.7$ where $Z_2$ monopoles are sufficiently
suppressed to extract continuum physics. We found that the connected
susceptibility subtraction used for the finite-temperature
transition at low net-baryon density does not work for
the  $\mu$-dependent additive renormalization of the chiral condensate
in the cold and dense matter. In the solid-state regime
around the half-filled lattice we have observed significant
differences between staggered and Wilson fermions, especially in the
behavior of the Polyakov loop.

More importantly, however, our prelimiary Goldstone spectrum outside
the bulk phase in Fig.~\ref{fig:gsmb1.7} demonstrates that the
continuum symmetry-breaking pattern of QC$_2$D is recovered. Comparing
the pion branches of Figs.~\ref{fig:goldstone_spectrum_beta1.5} and
\ref{fig:gsmb1.7} in the diquark-condensation phase we observe clear
indications that the correct antiunitary symmetries of QC$_2$D are
restored also with staggered quarks in the continuum limit. The most
likely reason is the realization of charge-conjugation invariance.


\begin{thebibliography}{10}

\bibitem{Hands:1999md}
  S.~Hands, J.~B. Kogut, M.-P. Lombardo and S.~E. Morrison,
  \href{http://dx.doi.org/10.1016/S0550-3213(99)00364-8}{\emph{Nucl. Phys.}
  {\bf B558} (1999) 327--346}.

\bibitem{Kogut:2001na}
  J.~B. Kogut, D.~K. Sinclair, S.~J. Hands and S.~E. Morrison,
  \href{http://dx.doi.org/10.1103/PhysRevD.64.094505}{\emph{Phys. Rev.} {\bf
      D64} (2001) 094505}.

\bibitem{Son:2000xc}
  D.~T. Son and M.~A. Stephanov,
  \href{http://dx.doi.org/10.1103/PhysRevLett.86.592}{\emph{Phys. Rev. Lett.}
    {\bf 86} (2001) 592--595}.

\bibitem{Kogut:2000ek}
  J.~B. Kogut, M.~A. Stephanov, D.~Toublan, J.~J.~M. Verbaarschot and
  A.~Zhitnitsky,\\
  \href{http://dx.doi.org/10.1016/S0550-3213(00)00242-X}{\emph{Nucl. Phys.}
  {\bf B582} (2000) 477--513}.

\bibitem{Kogut:2003ju}
  J.~B. Kogut, D.~Toublan and D.~K. Sinclair,
  \href{http://dx.doi.org/10.1103/PhysRevD.68.054507}{\emph{Phys. Rev.} {\bf
      D68} (2003) 054507}.

\bibitem{Kanazawa:2011tt}
  T.~Kanazawa, T.~Wettig and N.~Yamamoto,
  \href{http://dx.doi.org/10.1007/JHEP12(2011)007}{\emph{JHEP} {\bf 12} (2011)
    007}.

\bibitem{Strodthoff:2011tz}
  N.~Strodthoff, B.-J. Schaefer and L.~von Smekal,
  \href{http://dx.doi.org/10.1103/PhysRevD.85.074007}{\emph{Phys. Rev.} {\bf
      D85} (2012) 074007}.

\bibitem{Kamikado:2012bt}
  K.~Kamikado, N.~Strodthoff, L.~von Smekal and J.~Wambach,
  \href{http://dx.doi.org/10.1016/j.physletb.2012.11.055}{\emph{Phys. Lett.}
    {\bf B718} (2013) 1044--1053}.

\bibitem{Braguta:2016cpw}
  V.~V. Braguta, E.~M. Ilgenfritz, A.~{\relax Yu}. Kotov, A.~V. Molochkov and
  A.~A. Nikolaev,\\
  \href{http://dx.doi.org/10.1103/PhysRevD.94.114510}{\emph{Phys. Rev.} {\bf
      D94} (2016) 114510}.

\bibitem{Boz:2015ppa}
  T.~Boz, P.~Giudice, S.~Hands, J.-I. Skullerud and A.~G. Williams,
  \href{http://dx.doi.org/10.1063/1.4938682}{\emph{AIP Conf. Proc.} {\bf 1701}
  (2016) 060019}. 

\bibitem{Scior:2015vra}
  P.~Scior and L.~von Smekal,
  \href{http://dx.doi.org/10.1103/PhysRevD.92.094504}{\emph{Phys. Rev.} {\bf
      D92} (2015) 094504}.

\bibitem{Philipsen:2016wjt}
  O.~Philipsen, 
   for \emph{Confinement XII},
     Thessaloniki, Greece 2016, 
\newblock \href{https://arxiv.org/abs/1612.03400}{{\tt arXiv:1612.03400}}.

\bibitem{Clark2005}
  M.~A. Clark, \emph{{The Rational Hybrid Monte Carlo Algorithm}},
\newblock PhD thesis, Univ. of Edinburgh, 2005.

\bibitem{Halliday:1981tm}
  I.~G. Halliday and A.~Schwimmer,
  \href{http://dx.doi.org/10.1016/0370-2693(81)90630-4}{\emph{Phys.
  Lett.} {\bf B102} (1981) 337--340}.

\bibitem{Weisz:1983bn}
  P.~Weisz and R.~Wohlert,
  \href{http://dx.doi.org/10.1016/0550-3213(84)90563-7,
  10.1016/0550-3213(84)90543-1}{\emph{Nucl. Phys.} {\bf B236} (1984) 397}.

\bibitem{SchefflerDiss2015}
D.~Scheffler, \emph{{Two-Color Lattice QCD with Staggered Quarks}},
\newblock PhD thesis, TU Darmstadt, 2015.

\bibitem{Unger2010}
W.~Unger, \emph{{The Chiral Phase Transition of QCD with 2 + 1 Flavors}},
\newblock PhD thesis, Univ. Bielefeld, 2010.

\bibitem{Maas:2012wr}
  A.~Maas, L.~von Smekal, B.~Wellegehausen and A.~Wipf,
  \href{http://dx.doi.org/10.1103/PhysRevD.86.111901}{\emph{Phys. Rev.} {\bf
      D86} (2012) 111901}.

\bibitem{Wellegehausen:2013cya}
  B.~H. Wellegehausen, A.~Maas, A.~Wipf and L.~von Smekal,
  \href{http://dx.doi.org/10.1103/PhysRevD.89.056007}{\emph{Phys. Rev.} {\bf
      D89} (2014) 056007}.

\end{thebibliography}

\providecommand{\href}[2]{#2}\begingroup\raggedright\endgroup

\end{document}